\documentclass[12pt]{article}

\usepackage{graphicx}
\usepackage{amssymb}
\usepackage{amsmath}
\usepackage[dvipsnames,svgnames]{xcolor}
\usepackage{verbatim}
\usepackage{mathtools}

\usepackage{scicite}

\usepackage{times}

\newenvironment{sciabstract}{%
\begin{quote} \bf}
{\end{quote}}

\newcommand{\ajs}[1]{{{\color{red} #1}}}

\title{Acceleration is the Key to Drag Reduction in Turbulent Flow} 

\author
{Liuyang Ding$^{1}$, Lena Sabidussi$^{1}$, Brian C. Holloway$^{2}$, \\Marcus Hultmark$^{1}$ and Alexander J. Smits$^{1\ast}$ \\
\\
\normalsize{$^{1}$Mechanical and Aerospace Engineering, Princeton University,} \\
\normalsize{Princeton, NJ 08544, USA} \\
\normalsize{$^{2}$Intellectual Ventures, Bellevue, WA 98005, USA}\\
\\
\normalsize{$^\ast$To whom correspondence should be addressed; E-mail: asmits@princeton.edu}
}

\date{}

\begin{document} 

\baselineskip24pt

\maketitle

\begin{sciabstract}
A turbulent pipe flow experiment was conducted where the surface of the pipe was oscillated azimuthally over a wide range of frequencies, amplitudes and Reynolds number.  The drag was reduced by as much as 30\%.  Past work has suggested that the drag reduction scales with  the velocity amplitude of the motion, its period, or the Reynolds number.  Here, we find that the key parameter is simply the acceleration, which reduces the complexity of the phenomenon by two orders of magnitude.  This insight opens new potential avenues for reducing fuel consumption by large vehicles and for reducing energy costs in large piping systems.
\end{sciabstract}

In conditions encountered by airplanes, ships, wind turbines and pipelines, for example, turbulence generates skin-friction drag that constrains both speed and fuel efficiency. Even modest reductions in drag could immediately improve performance enough to yield significant economic and environmental benefits, such as improvements to the fuel efficiency of large vehicles and the handling capacity of industrial pipelines.
One of the most promising and most explored candidates for significant drag reduction is spanwise oscillation of surface elements, with drag reduction up to 50\% possible under some circumstances \cite{quadrio2009streamwise,ricco2021review}. 

Here we consider the case where the entire surface oscillates purely in time $t$ according to
\begin{equation}
    w_s(t) = A \sin(2 \pi t/T),
\end{equation}
in which $A$ is the amplitude of the spanwise velocity, $T$ is the period of the oscillation, and $x$ is the streamwise direction. We achieve this motion  in a pipe flow by oscillating the pipe in the azimuthal direction. 
The surface motion induces a spanwise velocity component in the flow very near the surface, well-described by a Stokes layer \cite{quadrio2011laminar}, and this momentum injection interferes with the turbulence production, specifically the streak-like structures found close to the surface, in such a way as to reduce drag.  
 
To relate our laboratory study to a full-scale application, we need to maintain dynamic similarity.  That is, we need to be able to scale the test result to the full-scale, and the crux of the current paper is to demonstrate that this scaling problem has an underlying simplicity which has not been appreciated before.

Consider that the local drag per unit area, averaged across the surface or “wall”, $\tau_w$, has a functional dependence $g_1$ given by:  
\begin{equation}
    \tau_w = g_1(\overline{U}, R,\rho, \nu,T,A),
\end{equation}
where $\overline{U}$ is is the bulk velocity, $R$ is the radius of the pipe, and $\rho$ and $\nu$ are the fluid density and kinematic viscosity, respectively. The drag reduction $DR$ is defined as the fractional decrease in $\tau_w$, and dimensional analysis then gives:
\begin{equation}
    DR = g_2(A^+, T^+, Re_\tau),
\label{DRfull}
\end{equation}
where $u_{\tau 0} = \sqrt{\tau_{w0}/\rho}$, and $Re_\tau = R u_{\tau 0}/\nu$ is the Reynolds number, which for full-scale applications is typically very large, that is, $Re_\tau \gg 10^3$. The subscript “0” denotes the quantities measured in the non-actuated case, that is, over the stationary surface.  The non-dimensional velocity $A^+ = A/ u_{\tau 0}$, and the non-dimensional period $T^+ = T u_{\tau 0}^2/\nu$. 
	
Flows with spanwise oscillating surfaces have been studied extensively in channel flows at low Reynolds numbers ($Re_\tau \leq 2000$) using Direct Numerical Simulations (DNS) (see, for example, \cite{quadrio2004critical,touber2012near,gatti2013performance,hurst2014effect,gatti2016reynolds,yao2019reynolds}). In this respect, Yao et al. \cite{yao2019reynolds} showed that at all Reynolds numbers investigated ($Re_\tau<2000$), the drag reduction rose with $T^+$ up to a peak value and then fell gradually at larger values of $T^+$. As the Reynolds number increased from 200 to 2000, the peak $DR$ decreased from a maximum about 35\% to about 23\%, and the optimal value of $T^+$ shifted from 100 to 80.  The scaling with $A^+$ was not investigated, since these computations were all done at a fixed value of $A^+ = 12$.

A recent computation for pipe flow with azimuthal oscillation reported 22.9\% DR at $Re_\tau=720$ ($A^+=10$, $T^+=100$) \cite{coxe2023reynolds} 
, and as far as the authors are aware there is only one previous experiment reported in the literature \cite{choi1998drag, choi2001mechanism,choi2002drag}. 
The drag reduction was measured at $Re_\tau$ = 649 and 995 over a relatively large range of $A^+$ and $T^+$ values, and it displayed a very similar behavior to that found using DNS in a channel at comparable Reynolds numbers.  That is, the maximum $DR$ reached approximately 25\% at $T^+ \approx 100$, maintaining this value down to $T^+ \approx 50$, the lowest value explored in these experiments. The authors concluded, without much justification, that the results scaled on $T^+$.

All these studies were for $Re_\tau \leq 2000$.  To examine the behavior at higher Reynolds numbers, and to investigate more fully the scaling on $A^+$ and $T^+$, we cover Reynolds numbers from 1341 to 6851, and use a novel strategy of controlling the water temperature to vary $A^+$, $T^+$ and $Re_\tau$ independently.  As seen from Fig. 1, neither $A^+$ nor $T^+$ collapse our data.  Although some of our results in these scalings agree with the data of Choi \& Graham \cite{choi1998drag}, other results do not. This is particularly true for the dependence on $T^+$.  

A more compelling result is revealed when we plot the data in terms of the non-dimensional acceleration $a^+ = A^+/T^+ = A \nu /(T u_{\tau 0}^3)$, as in Fig. 2.   Immediately, we see a convincing collapse of our data (Fig. 2a), and our data now also agrees well with that of Choi \& Graham \cite{choi1998drag}, at least for $T^+ > 150$ (Fig. 2b).  This is in marked contrast to previous scaling based on $A^+$ or $T^+$, or on other parameters such as the influence range of the Stokes layer   
\cite{choi2002drag, quadrio2004critical}, as shown in the Supplementary Materials.
  
We see that, for the range of parameters investigated here,
\begin{equation}
    DR = g(a^+).
\label{DRfinal}
\end{equation}
Therefore it is the applied force that is the controlling parameter in drag reduction, which in hindsight seems evident. Crucially, compared to the initial problem described by equation~\ref{DRfull}, this result represents a reduction in complexity of two orders of magnitude.  

The results for values of $T^+ > 150$ are of particular interest because the power required to move the fluid scales approximately as $(T)^{-2.5}$ \cite{quadrio2011laminar} and so there are great benefits to operating at low frequencies. Such low-frequency, large-period actuation appears to be the only pathway to energy-efficient drag reduction at high Reynolds number \cite{marusic2021nature, Chandran2022part2}.

What about cases where the actuation includes a streamwise traveling wave component, as in $ w_s(x,t) = A \sin(2 \pi x/\lambda- 2 \pi t/T)$, where $\lambda$ is the wavelength of the traveling wave?  Here, the initial problem has an additional non-dimensional parameter $\lambda^+=\lambda u_{\tau 0}/\nu$.  The only high Reynolds number data available are those from boundary layer experiments with $Re_\tau \ge 4500$ \cite{marusic2021nature, Chandran2022part2}.  As is evident from Fig. 3, for $T^+ >150$ the acceleration is still the unifying scaling parameter, independent of the wavelength and Reynolds number. At low values of $a^+$, however, these experiments show higher DR than the pipe flow cases, for all wavelengths investigated. In contrast, at higher values of the acceleration, the boundary layer drag reduction appears to asymptote to a value that lies below the corresponding pipe flow values.  These observations could indicate important differences in $DR$ behavior for pipes compared to boundary layers, or they could indicate that traveling wave motion has an important effect over this range of wavelengths.  Further work is required to settle these questions.  Nevertheless, the central role of the acceleration in scaling the drag reduction is evident.

What about cases with non-sinusoidal oscillations?  Since the acceleration is the key parameter, it may well be advantageous to use non-sinusoidal waveforms where the acceleration can be maximized.
As noted by \cite{touber2012near} and \cite{deshpande2023near}, the attenuation of the near-wall streaks only occurs during certain `active' phases of the wall-oscillation cycle when the orientation of the shear strain vector changes rapidly. 
The shear vector tends to change very little (i.e. it `lingers') in other `inactive' phases of the cycle, which permits (re-)generation of the near-wall streaks, and consequently, recovery to a relatively higher drag state.  Controlling the acceleration amplitude and phase may be the key to reducing this dwell time, and therefore non-sinusoidal waveforms might achieve drag reduction even more efficiently than sinusoidal waveforms.

\section*{Acknowledgments}
The research was funded by the Deep Science Fund of Intellectual Ventures.

\section*{Supplementary materials}

{\bf Materials and Methods}

The experiments were conducted in a recirculating water pipe facility with an inner diameter ($2R = D$) of 38.1 mm. Transverse momentum injection was implemented by oscillating the pipe around its longitudinal axis (see Fig. S1). The oscillating pipe section is of length 1.21 m connected to a crank-slider mechanism driven by a motor via a T-slot that allows for changes in the oscillation amplitude. The slider translates linear motion to azimuthal oscillation of the pipe via a timing belt. The length ratio of the crank to the rod connecting the crank to the slider is less than 0.05, so that the oscillation closely approximates a sinusoidal motion. Oscillation azimuthal amplitudes ($d$) up to 12.8 mm and oscillation frequencies ($f = 1/T$) up to 20 Hz were tested. The test section was placed $100D$ downstream of the inlet to ensure fully developed flow at the entrance to the test section. Experiments were performed at bulk velocities ranging from 1.1 to 4.2 m/s, and the water temperature was adjusted from ambient values up to 57$^\circ$C. The Reynolds number was varied from about 1000 to about 7000. Table S1 summarizes our various test parameters, together with those from the oscillating pipe experiment by Choi \& Graham \cite{choi1998drag} . 

To measure the drag reduction, the friction factor was found by measuring the time-averaged pressure drop using a pair of pressure taps located 135 mm upstream and downstream of the oscillating pipe section. A differential pressure sensor, Validyne DP103, was used, and it was regularly calibrated throughout our measurement campaign to ensure an accuracy within 1$-$2\% as compared to the friction factor correlation reported by McKeon et al. \cite{mckeon2004friction}. Pressure difference data were adjusted by subtracting the pressure drop in the upstream and downstream stationary sections, assuming drag alteration occurred entirely within the oscillating section. $DR$ was then computed as the percentage difference of the friction factor, using the results from McKeon et al. \cite{mckeon2004friction} for the non-oscillating case at matching $Re_\tau$ values. For each combination of $[A^+, \, T^+, \, Re_\tau]$, we acquired a total of 2000$-$3000 samples, in one or two trials, at a sampling rate of about 5 Hz. This sampling frequency ensured that pressure measurement was not phase-locked with respect to pipe oscillation in all datasets. 

The uncertainty of $DR$ was considered to have two contributions. First, the uncertainty due to turbulent fluctuation and random measurement noise combined may be approximated as $\Delta p_\mathrm{rms}/\sqrt{N_s}$, where $\Delta p_\mathrm{rms}$ is the root-mean-square (rms) pressure drop and $N_s$ is the number of samples. This yielded a maximum uncertainty of 0.2\%. Second, the pressure sensor had a bias error of within 1$-$2\% when compared to the results in \cite{mckeon2004friction}, as described earlier. The latter therefore dominates the total uncertainty, and we thus claim a $\pm 2\%$ uncertainty (i.e. $DR\pm2\%$) for all of our $DR$ data presented in Fig. 1 and 2.

\vspace{6pt}

\noindent {\bf Supplementary Text}

Quadrio \& Ricco \cite{quadrio2004critical} proposed a scaling using the parameter $S^+ = a_m^+l^+/A_m^+$, where $A_m^+$ and $a_m^+$ are the maximum spanwise velocity and acceleration, respectively, during a cycle, and $l^+$ is the penetration depth of the Stokes layer.  The $DR$ data shown to collapse with the non-dimensional acceleration $a^+$ in Fig. 2b are replotted in Fig. S2 against $S^+$. Clearly, no convincing collapse is observed.  Similar results were obtained using the scaling parameter $V_c^+$ suggested by Choi et al. \cite{choi2002drag}, where $V_c^+=a_5^+ y_d^+/(A^+Re_\tau^{0.2})$, $y_d^+$ represents the influence range of the Stokes layer, and $a_5^+$ is a measure of the acceleration of the Stokes layer at $y^+=5$.

\clearpage

\begin{figure}
    \includegraphics[width=0.49\textwidth]{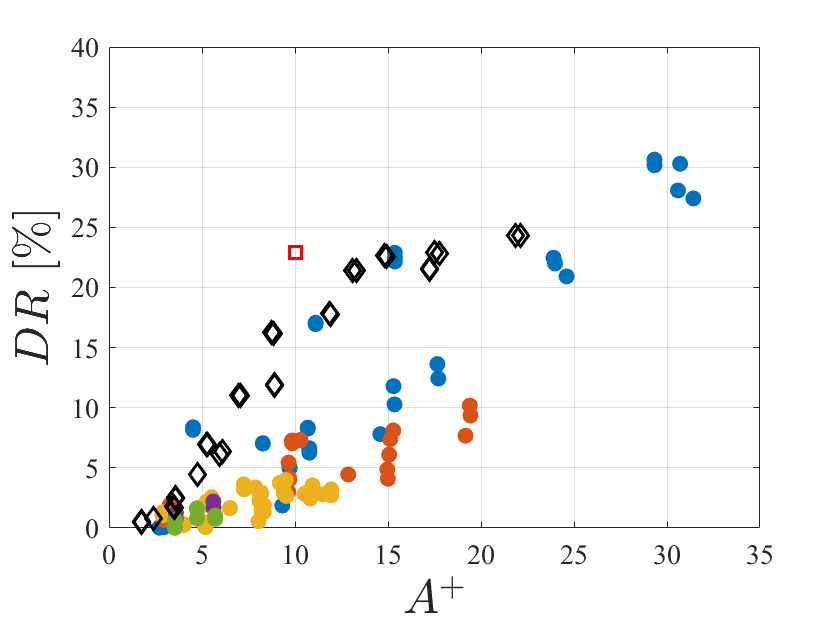}
    \includegraphics[width=0.49\textwidth]{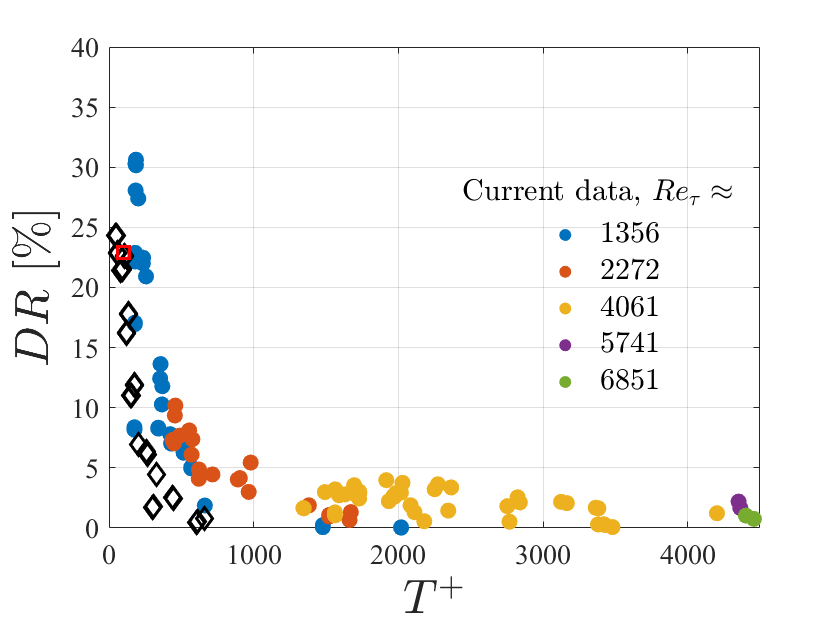}
\end{figure}

\noindent {\bf Fig. 1.} Drag reduction versus (a) non-dimensional velocity amplitude of oscillation $A^+$; (b) non-dimensional period of oscillation $T^+$.  Filled circles, current data color coded by $Re_\tau$ (see legend); {\Large $\diamond$}, pipe flow data  \cite{choi1998drag}; \ajs{{\footnotesize $\square$}}, pipe flow DNS  \cite{coxe2023reynolds}. 

\clearpage

\begin{figure}
    \includegraphics[width=0.49\textwidth]{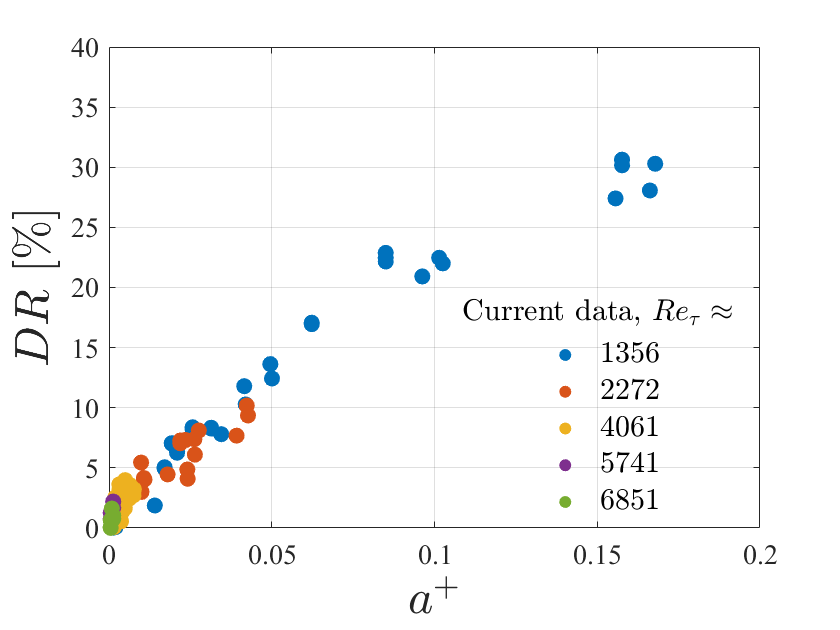}
    \includegraphics[width=0.49\textwidth]{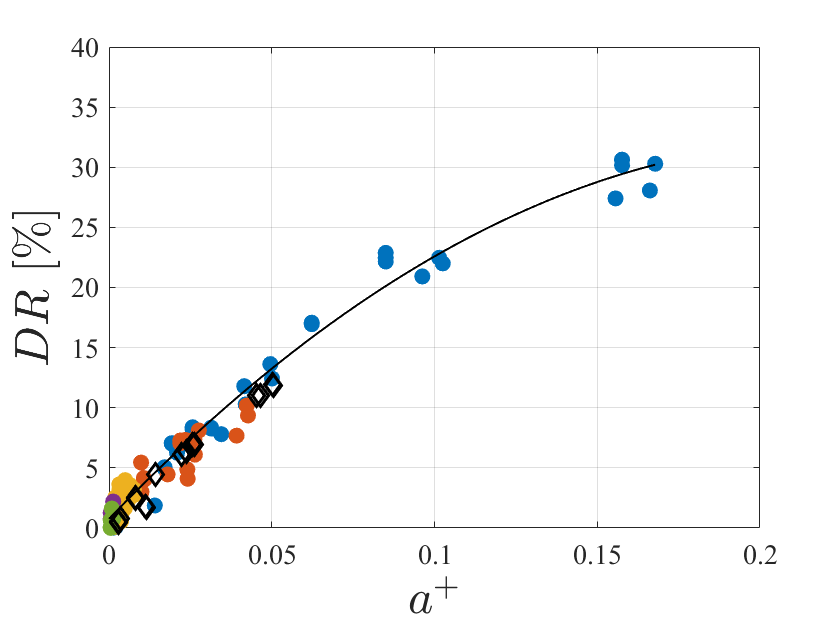}
\end{figure}

\noindent {\bf Fig. 2.} Drag reduction versus non-dimensional acceleration amplitude $a^+$. (a)  All current data.  (b) All data for for $T^+ > 150$.  Filled circles, current data color coded by $Re_\tau$ (see legend); {\Large $\diamond$}, pipe flow data \cite{choi1998drag}.  Line for guidance only. 

\clearpage

\begin{figure}
\centering
\includegraphics[width=0.50\textwidth]{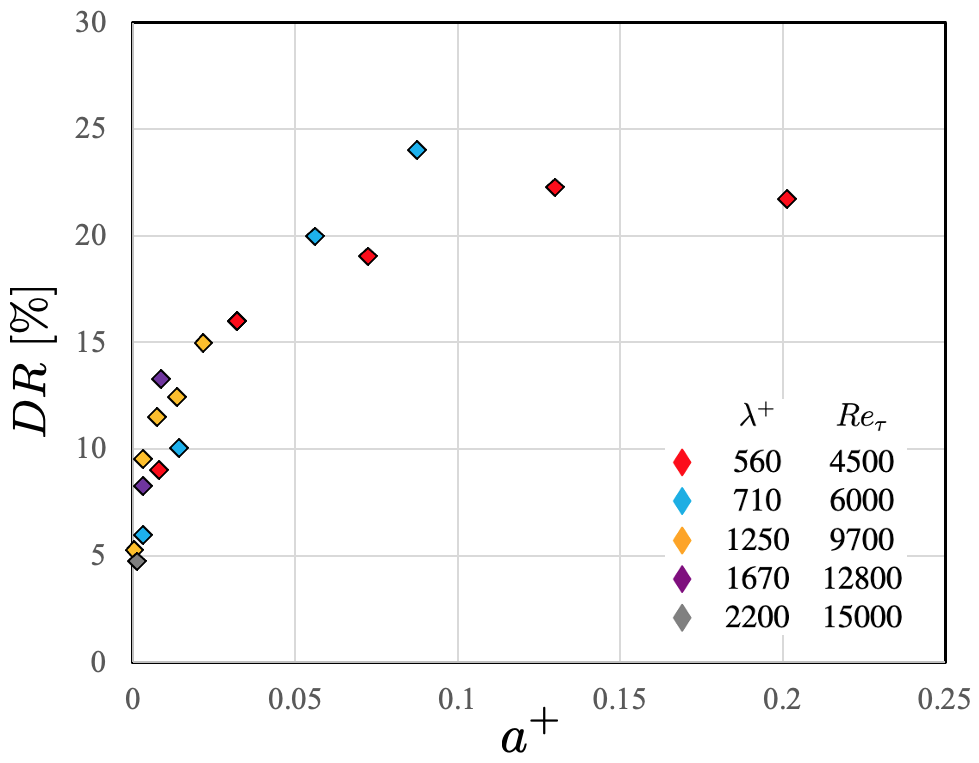}
\end{figure}

\noindent {\bf Fig. 3.} Boundary layer data with  streamwise traveling wave for $T^+ >150$.  
Data from \cite{Chandran2022part2}.

\clearpage

\begin{figure}
\centering
\includegraphics[width=0.75\textwidth]{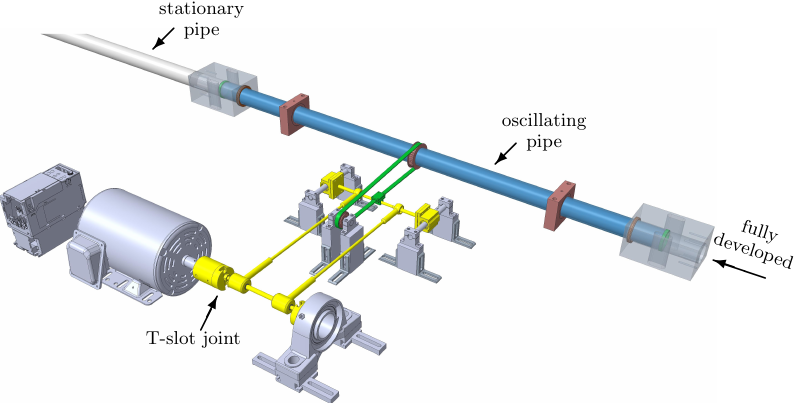}
\end{figure}

\noindent {\bf Fig. S1.} Model of the oscillating pipe driven by a crank-slider mechanism (shown in yellow) via a timing belt (shown in green). The crank is connected to the motor via a T-slot permitting adjustable amplitudes.

\clearpage

\begin{figure}
\centering
\includegraphics[width=0.50\textwidth]{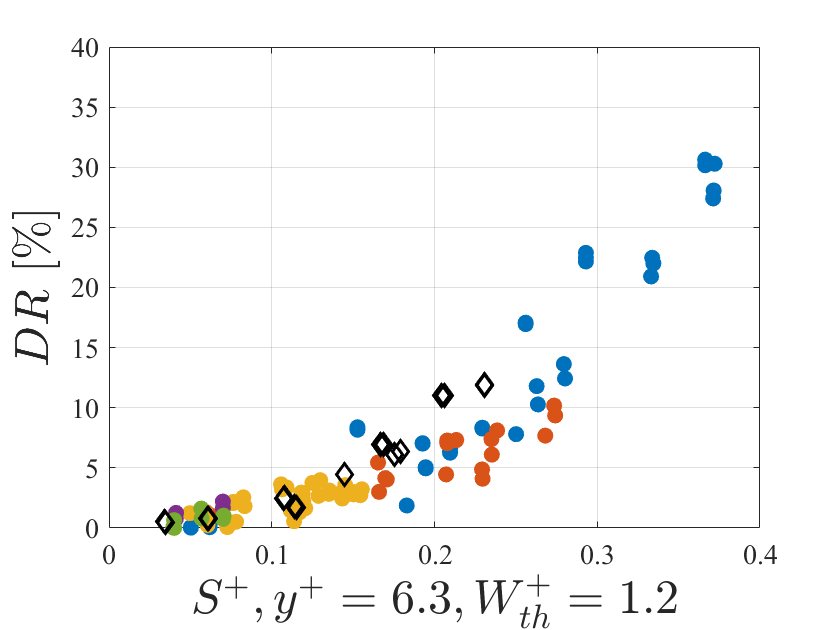}
\end{figure}

\noindent {\bf Fig. S2.} The $DR$ data shown in Fig. 2b are replotted against the scaling parameter $S^+$ proposed by Quadrio \& Ricco \cite{quadrio2004critical}. See Fig. 1 for legends. 

\clearpage

\begin{tabular}{lcccccc}
\hline \\
 &  \hspace{2mm} $Re_\tau$ \hspace{2mm} & \hspace{3mm} $A^+ $ \hspace{3mm} & \hspace{3mm}   $T^+ $  \hspace{3mm} &  \hspace{3mm} $DR (\%) $  \hspace{3mm}   \\[2mm]

Experiment                                
&~649 &1.76$-$22.1  &608$-$48 &  0.5$-$24 \\
(Choi \& Graham 1998)  &~995 &2.37$-$17.2  &663$-$91 &  0.8$-$22 \\[3mm]

DNS                                
(Coxe et al. 2023)  &~720 & 10  & 100 &  22.9 \\[3mm]

Experiment                   & 1356 $\pm$ 52 & 2.72$-$31.4 & 2019$-$176 &  0.0$-$30.6 \\
(current data)         & 2272 $\pm$ 318 & 2.96$-$19.4 & 1672$-$439 &  0.7$-$10.2 \\
                       & 4061 $\pm$ 380 & 2.88$-$11.95 & 4203$-$1343 &  0.1$-$4.0 \\
                       & 5741 $\pm$ 23 & 3.59$-$5.62 & 6857$-$4352 &  0.8$-$2.2 \\
                       & 6851 $\pm$ 75 & 3.54$-$5.71 & 7290$-$4403 & 0.0$-$1.6 \\[3mm]
\hline \\

\end{tabular} 

\noindent {\bf Table S1.} Test parameters for oscillating pipe cases.  Only the `fully turbulent' DNS case at $Re_\tau=720$ was used here.

\end{document}